\documentclass[
preprint,
 amsmath,amssymb,
 aps,
]{revtex4-2}

\usepackage{cmbright}
\DeclareFontShape{OT1}{cmss}{m}{it}{<->ssub*cmss/m/sl}{}

\newcommand{\diff}{\mathop{}\!\mathrm{d}}
\usepackage{graphicx}% Include figure files
\usepackage{dcolumn}% Align table columns on decimal point
\usepackage{bm}% bold math
\usepackage{xcolor}
\usepackage{comment}
\begin{document}

\preprint{APS/123-QED}

\title{Emergent intelligence of buckling-driven elasto-active structures}% Force line breaks with \\

\author{Yuchen Xi $^{1}$, Trevor J. Jones$^{1}$, Richard Huang$^{1}$, Tom Marzin$^{1}$, P.-T. Brun$^{1}$ }

\affiliation{$^1$Department of Chemical and Biological Engineering, Princeton University, Princeton, New Jersey 08540, USA
}

\author{}
 \altaffiliation{}%Lines break automatically or can be forced with \\
 \email{pbrun@princeton.edu}

\date{\today}% It is always \today, today,

\begin{abstract}
Active systems of self-propelled agents, e.g., birds, fish, and bacteria, can organize their collective motion into myriad autonomous behaviors ~\cite{marchetti2013hydrodynamics}. Ubiquitous in nature and across length scales, such phenomena are also amenable to artificial settings, e.g., where brainless self-propelled robots orchestrate their movements into spatio-temportal patterns via the application of external cues or when confined within flexible boundaries~\cite{michelin2023self,giomi_swarming_2013,boudet_collections_2021,deblais_boundaries_2018,scholz_rotating_2018,sepulveda_bioinspired_2021,li_programming_2021,boudet_effective_2022,dibari_using_2022}. Very much like their natural counterparts, these approaches typically require many units to initiate collective motion such that controlling the ensuing dynamics is challenging. Here, we demonstrate a novel yet simple mechanism that leverages nonlinear elasticity to tame near-diffusive motile particles in forming structures capable of directed motion and other emergent intelligent behaviors. Our elasto-active system comprises two centimeter-sized self-propelled microbots connected with elastic beams. These microbots exert forces that suffice to buckle the beam and set the structure in motion. We first rationalize the physics of the interaction between the beam and the microbots. Then we use reduced order models to predict the interactions of our elasto-active structure with boundaries, e.g., walls and constrictions, and demonstrate how they can exhibit intelligent behaviors such as maze navigation. The findings are relevant to designing intelligent materials or soft robots capable of autonomous space exploration, adaptation, and interaction with the surrounding environment.
\end{abstract}

\maketitle
%\tableofcontents
\section{Introduction}
The study of active matter, living or inert, focuses on understanding the mechanical and statistical properties of systems comprising elements capable of converting energy into movement. The field is particularly interested in identifying the principles governing the emergence of self-organized spatio-temporal patterns on scales larger than individual motile units. Examples range from liquid-crystalline order in bacterial flocks to polar order in a school of fish\cite{marchetti2013hydrodynamics}. While common in nature, active matter systems are also amenable to artificial laboratory systems\cite{michelin2023self}. Exploring model experimental systems allows a careful investigation of the inner workings of active matter, particularly identifying the onset of collective behaviors and rationalizing pattern formation within bulk ensembles of active particles. Historically, the field has focused heavily on fluids and fluid-like systems\cite{marchetti2013hydrodynamics}, making active elastic systems comparatively less explored\cite{fruchart2023odd}. 

In recent years, self-propelled microbots, e.g., Hexbug Nano\textregistered\cite{hexbug}, have been identified as a tunable and reliable means for developing active structures, e.g., oscillatory tails\cite{zheng_self-oscillation_2023} and active elastic solids\cite{baconnier2022selective}. The motion of individual microbots is understood as vibrating masses whose frictional contacts cause propulsion\cite{cicconofri_motility_2015,ben_zion_morphological_2023,kim_forward_2020}, which can be modeled as self-propelled particles that follow Langevin dynamics on timescales much longer than the vibration frequency of their body. This approach allows for the modeling of microbots dynamics in confined geometries\cite{giomi_swarming_2013,leoni_surfing_2020} or in a harmonic trap\cite{dauchot2019dynamics}. The collective behavior of such microbot systems has received particular attention \cite{giomi_swarming_2013,boudet_collections_2021,deblais_boundaries_2018,scholz_rotating_2018}: in bounded and crowded environments, these microbots can display a gas-like behavior\cite{boudet_effective_2022,dibari_using_2022} or cluster around the edges of boundaries\cite{deblais_boundaries_2018,giomi_swarming_2013,boudet_collections_2021}. In addition, external cues such as light and magnets can be used to control such robotic swarms, e.g., to form clusters or direct movements\cite{sepulveda_bioinspired_2021,li_programming_2021}. However, such methodologies are still in their infancy so that finding means to effectively and efficiently control such microbot systems remains an ongoing effort essential to developing robotic matter capable of achieving autonomous, predictive, and tunable motions.

\begin{figure}[!ht]
    \includegraphics[width=0.9\textwidth]{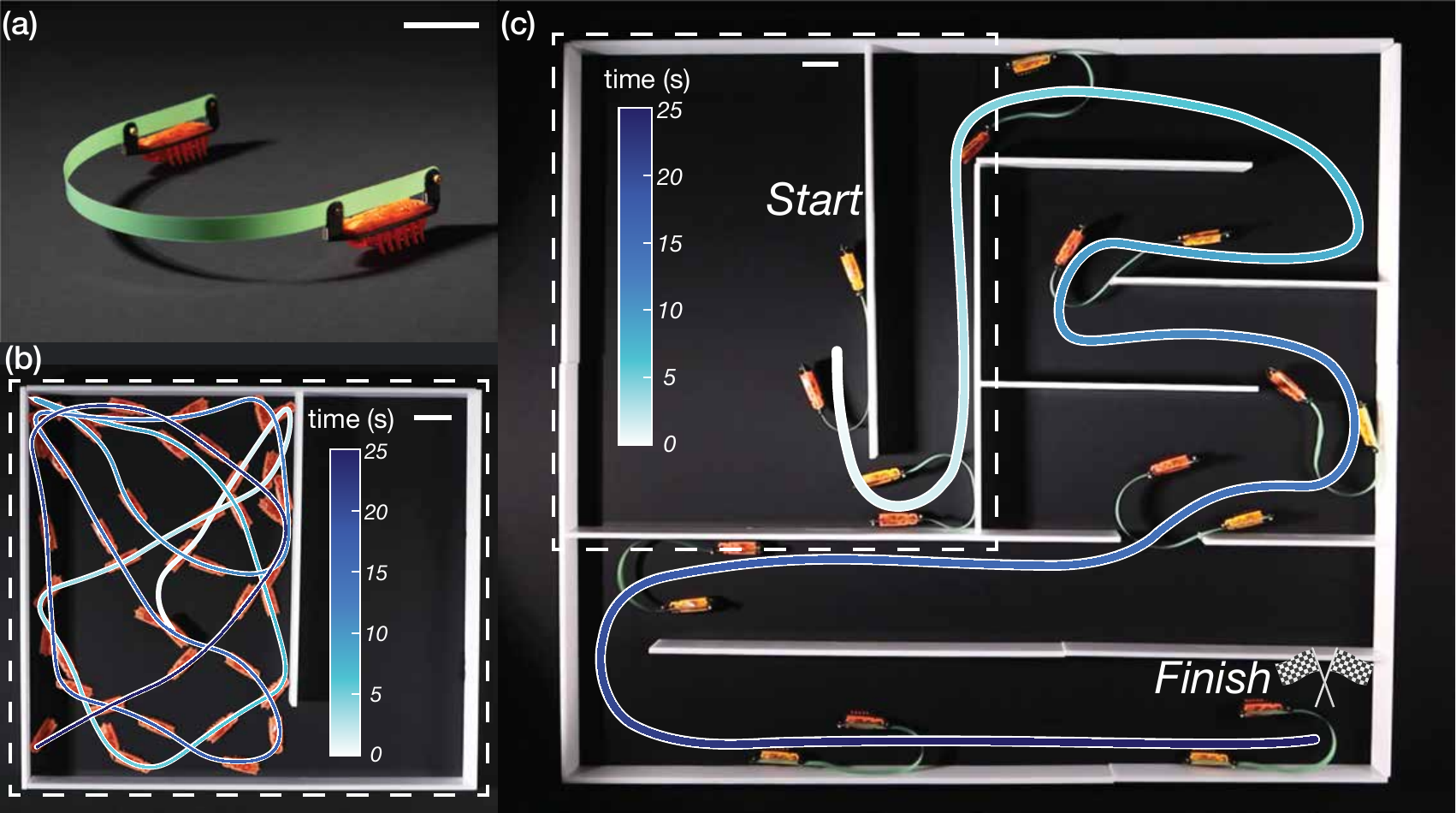}
    \caption{\textbf{From mindless particles to emergent intelligence} (a) Photograph of the bucklebot, showing two microbots connected by a thin polyester beam. (b) Individual microbot trajectory in a confined space. (c) A bucklebot efficiently navigates a maze within 25 seconds. The dashed area in (c) matches the space shown in (b). (all scale bars are 50 mm in length, and trajectories are color-coded by time).}
    \label{fig1}
\end{figure}

Here, we introduce a new form of autonomous physical behavior by coupling active particles with nonlinear elasticity. Fig. \ref{fig1}(a) illustrates our approach involving two self-propelled microbots connected by an elastic polyester beam. We operate in a regime where the active force exerted by the microbots is sufficient to buckle the connecting beam, thereby aligning the microbors and allowing this contraption, coined \textit{bucklebot}, to move across a flat substrate. While individual microbots remain trapped in a confined space for prolonged periods (Fig. \ref{fig1}(b)), a bucklebot manages to solve a maze efficiently, as evident in Fig. \ref{fig1}(c) and Movie S1. Combining experiments and theory, we elucidate the physics governing the dynamics of these bucklebots. We then explore the interaction of bucklebots with physical boundaries, e.g., plane walls and narrow constrictions. Finally, we leverage these quantitative results to elucidate how bucklebots can develop emergent intelligent behaviors such as solving a maze, probing a path, or organizing disperse particles.

\section{Results}

\subsection{bucklebot characterization}
\begin{figure}[!ht]

    \includegraphics[width=0.9\textwidth]{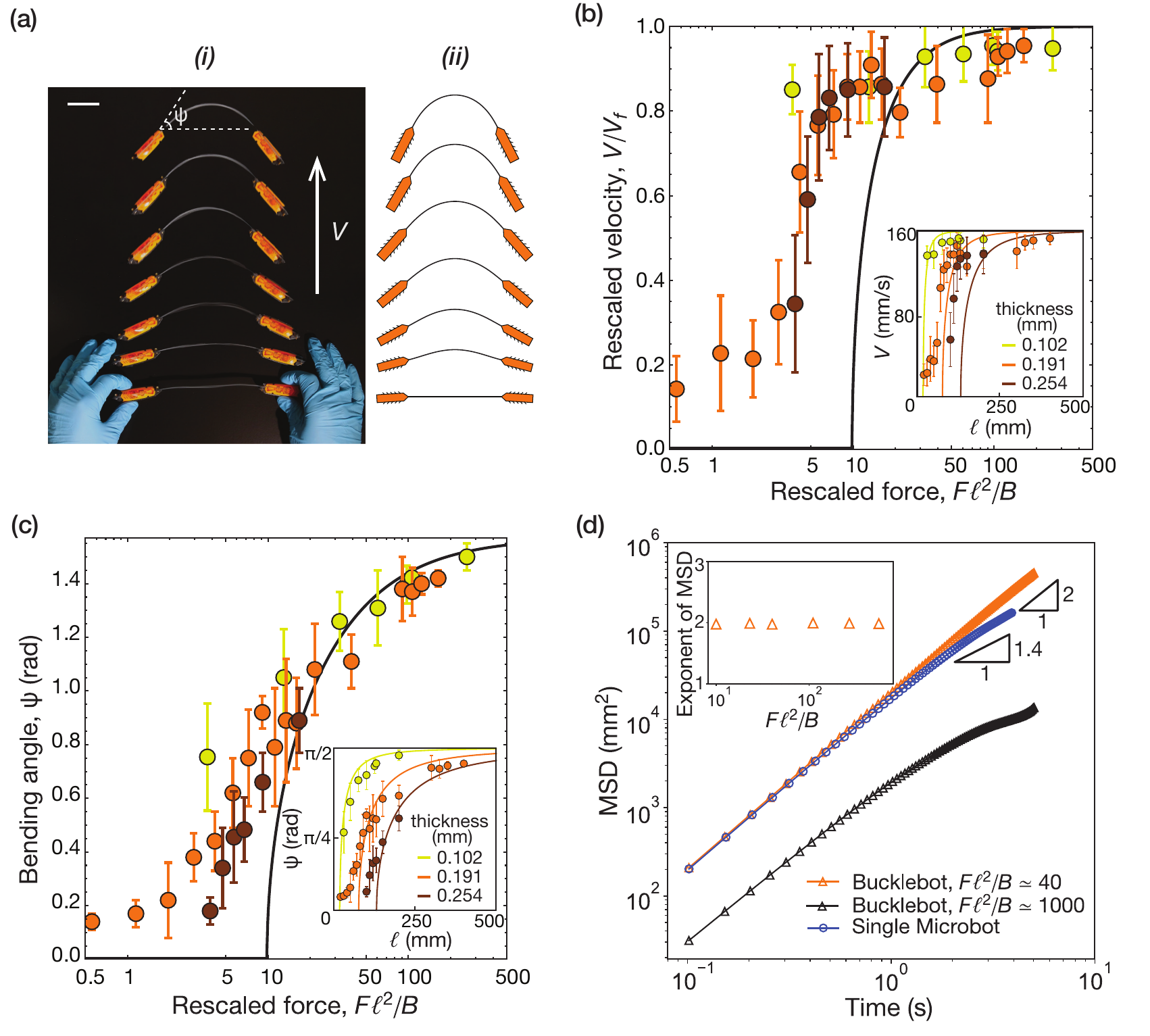}
    \caption{\textbf{Dynamics and characterization of bucklebots:} (a) (i): Timelapse of a bucklebot with $\Delta t=0.1 s$ (scale bar=50 mm); (ii) Bucklebot dynamics obtained by integrating our model  (SI Section A-C).  (b) Rescaled velocity $V/V_f$ and (c) bending angle $\psi$ versus rescaled force $F\ell^2/B$. The black line represents the predicted steady-state solution with $\lambda \simeq 0$ (SI Section D). Inset: bucklebot velocity $V$ and bending angle $\psi$ plotted against beam length $\ell$ for three beam thicknesses. Lines represent the steady-state solutions of Eqns. \ref{eq:beam_force_non}-\ref{eq:beam_mom_non} in SI (d) Log-log plot of the mean squared displacement (MSD) versus time for a single microbot (blue), a bucklebot with $F\ell^2/B \simeq$ 40 (orange), and a bucklebot with $F\ell^2/B \simeq$ 1000 (black). Inset: MSD exponents for bucklebots versus $F\ell^2/B$.}
    \label{fig2}
\end{figure}
Figure \ref{fig2} summarizes the main results pertaining to bucklebots evolving in free space. In Fig. \ref{fig2}(a), we show the onset of their motion. Namely, when released, the microbots progressively bend the beam that connects them before assuming a final steady-state configuration characterized by a bending angle $\psi$ and a steady-state velocity $V$, reached after nearly a second. In Fig. \ref{fig2}(b)-(c), we show the variation of these observables when the length and thickness of the beam are varied. For relatively short and thick (thus stiff) beams, the angle $\psi$ remains close to zero, and the structure barely moves. For longer and thinner (thus soft) beams, the force exerted by the microbots is sufficient to buckle the beam, increasing $\psi$ until the limit value of $\pi/2$ is approached. At this point, the microbots are parallel, facing the same direction and moving at a speed close to their free velocity $V_f$. The value of $V_f$ is typically related to the force exerted by the microbots and the friction between the structure and the substrate, $V_f = F /\gamma$ where $F$ is the microbot force and $\gamma$ is the effective drag coefficient acting on the microbots\cite{Weber2013}. 

We recast our experimental data in dimensionless form using $V_f$ as our speed gauge and $B/\ell^2$ as the force gauge that captures the beam resistance to bending, where $B$ is the bending stiffness and $\ell$ is the length of the beam. In Fig. \ref{fig2}(b)-(c), we show that our data collapse to a single master curve, confirming the relevance of rescaled force $F\ell^2/B$ in predicting the system behavior. Our experiments show a non-zero velocity and bending angle even for small values of $F\ell^2/B$. While the microbots cannot buckle the beam, the bucklebot slides or rotates slowly due to the vibrations from the motor. As $F\ell^2/B$ increases, both $\psi$ and $V$ increase until they reach a plateau around $F \ell^2/B \simeq 50$. Overall, this transition and the overall variation in geometry and speed are favorably recovered by our model, which is obtained by combining the Kirchhoff equations for elastic beams with a force and moment balance for microbots (See SI Eqn. \ref{eq:Kirchhoff}-\ref{eqs:hex}). The difference between experiment and theory is attributed to a finite size effect: the microbots are not point masses, so a third dimensionless number $\lambda=L/\ell$ is introduced to describe their length relative to that of the beam. In the limit case where $\lambda \simeq 0$, the transition between static and translation occurs at $F\ell^2/B \simeq 10$, in agreement with Euler's critical load for column ends with hinge-hinge boundary conditions\cite{audoly2000elasticity}. In contrast, for larger values of $\lambda$, the microbots exert higher lever-arm torques onto the beam, thereby diminishing the critical buckling load (See SI Section E). 

Having understood the shape and instantaneous velocity of our bucklebots we move to describe their long-term behavior. In Fig.\ref{fig2}(d) we calculate their mean square displacement MSD = $\langle |\boldsymbol{r}(t)-\boldsymbol{r}(0)|^2\rangle$, where $\boldsymbol{r}(t)$ is the position vector at time $t$, and $\langle\cdot\rangle $ denotes the average value over of all recorded trajectories. In Fig. \ref{fig2}(d), we plot the MSD of two bucklebots with rescaled forces of 40 and 1000 together with that of single microbots. Single microbots show diffusive-like behavior resembling a noisy walker\cite{FODOR2018106} with a reorientation time $\tau \simeq 1.3$ s and long-term MSD $\propto t^{1.4}$. In contrast, the bucklebots with $F\ell^2/B \simeq 40$ translate ballistically (MSD $\propto t^2$) in the range of the time ($>$$5\tau$) we probed. Bucklebots achieve persistent directed motion despite the direction changes typically observed in each unit and their inevitable differences. This result remains true for $10<F\ell^2/B<600$, where similar behaviors are observed (see inset). However, past this upper limit, bucklebots demonstrate slower movement and cover two orders of magnitude smaller areas throughout the measurement, as evident from the orange line. In such high-force regimes, the beam's internal resistance to bending is negligible and thus insufficient to align with the motions of the microbots. The microbots tend to buckle the beam to its second (and higher) buckling modes, so the bucklebot rotates while slowly translating (See Movie S2). In the following, we focus on bucklebots with $F\ell^2/B$ within the range from 10 to 600 and probe their interactions with boundaries.

\begin{figure}[!ht]

    \includegraphics[width=0.85\textwidth]{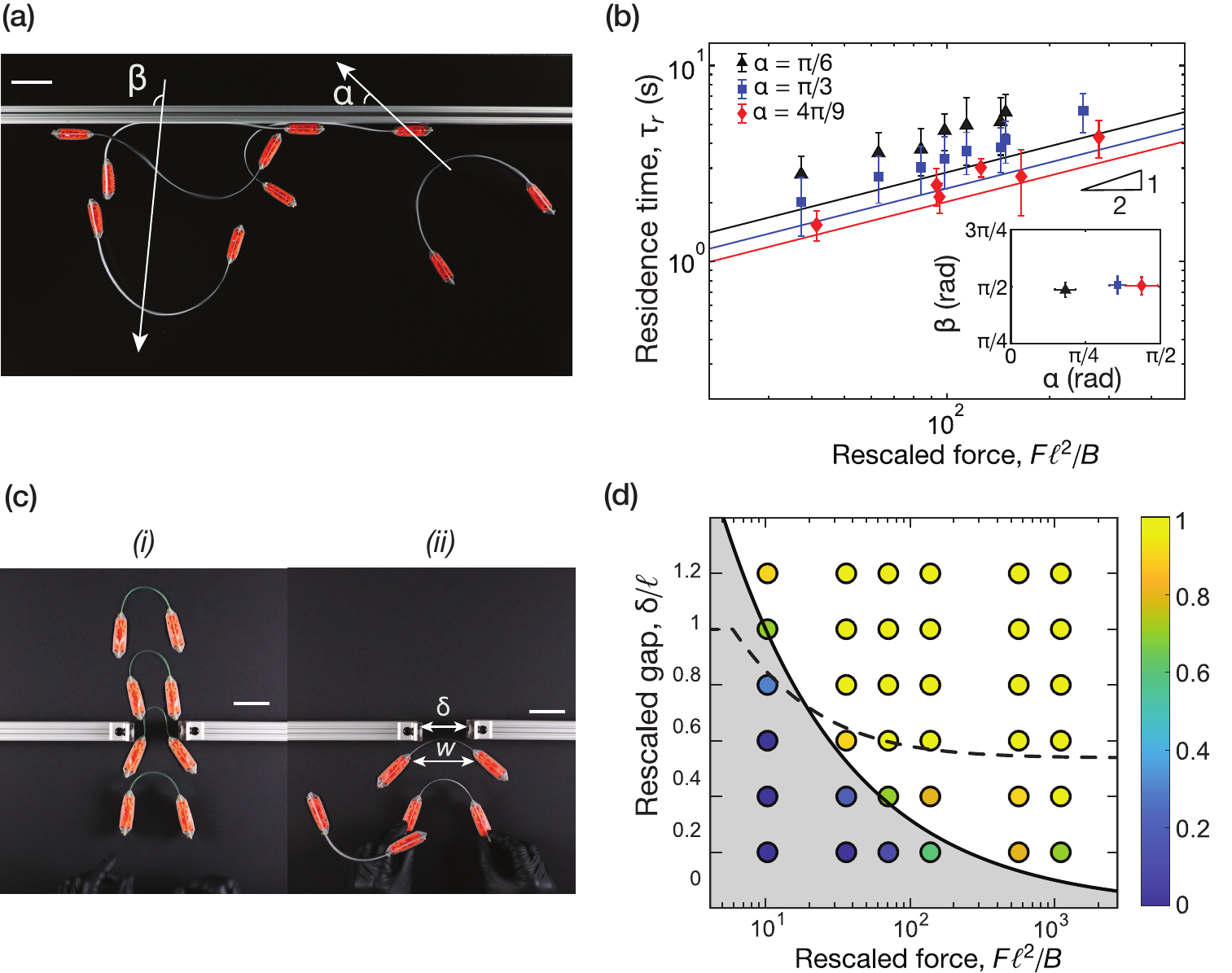}
    \caption{\textbf{Bucklebots interacting with boundaries} (a) Overalyed photographs of bucklebot with $F\ell^2/B \simeq 60$ approaching a flat wall with angle $\alpha$, following the wall for some time $\tau_r$, and bouncing off with a reflection angle $\beta$ (scale bar=50 mm). (b) Residence time, $\tau_r$ versus $F\ell^2/B$ for three sets of $\alpha$. Markers represent experiments (triangles: $\alpha=\pi/6$, squares: $\alpha=\pi/3$, diamonds: $\alpha=4\pi/9$). Lines are the predictions from the self-oscillation model (see SI Section G). The error bars represent the standard deviation of $\tau_r$ for each bucklebot. The inset shows $\beta$ versus $\alpha$. (c) Snapshots of passage through a slit of width $\delta=6$ cm. (i): a bucklebot with $F\ell^2/B \simeq 140$ and (ii) with $F\ell^2/B \simeq 13$ (scale bar=50 mm). (d) Success of passage through the slits over ten launches, shown as a function of rescaled gap size $\delta/\ell$ and $F\ell^2/B$.  The experiment data is color-coded by the success rate of passage, as shown by the right color bar. The dashed line indicates the equilibrium width of the free bucklebot (see SI Section D), and the solid line corresponds to our model (SI Eqn. \ref{eq:hinge_hinge3}). The shaded gray area is our prediction for the region where the bucklebots are expected to bounce off from the slit.}
    \label{fig3}
\end{figure}

We first turn our attention to the interaction of a bucklebot with a plane boundary (see Fig. \ref{fig3}(a)). The bucklebot approaches the wall with an angle $\alpha$ and is found to follow the wall for some residence time $\tau_r$ before reflecting off with an angle $\beta$. In Fig. \ref{fig3}(b), we find that the reflection angles $\beta$ are consistently around $\pi/2$, irrespective of the value of $\alpha$. However, the residence time $\tau_r$ increases as $\alpha$ decreases. Shallower approaches stay longer along the wall than a direct hit. Additionally, we find that $\tau_r \propto \sqrt{F\ell^2/B}$.  To rationalize such a scaling law, we observe that the microbot in contact with the wall is typically slower than the other one, presumably because of the added friction. As such, the faster outer microbot overtakes its slower counterpart and forces the beam to snap (See Movie S3). Inspired by such behavior, we introduce the limit case scenario, where one single microbot is attached to an elastic beam clamped on one end. We model the ensuing oscillatory dynamics (See SI Eqns. \ref{eq:tail_bc1}-\ref{eq:tail_bc2}) and recover the scaling law observed in experiments, as indicated by the solid lines in Fig. \ref{fig3}(b). Our model underpredicts our data since, in our experiment, the bucklebot at the wall is not clamped but instead slides, thereby delaying the beam's oscillation. 

Next, we turn to study the passage of a bucklebot through constrictions. Figure \ref{fig3}(c)(i) illustrates the bucklebot ability to deform and pass a tight slit with opening $\delta<w$, with $w$ the bucklebot width. If the beam is too stiff or the slit is too small, the buckle-bot will bounce off the constriction (See Figure \ref{fig3}(c)(ii)). Those results are formalized in  Figure \ref{fig3}(d), where we report the probability of successful passage as a function of the gap size rescaled by the beam length, $\delta/\ell$, and the 
rescaled force, $F\ell^2/B$. As evident from the figure, larger slits, and larger forces correlate with a higher probability of successful passage.
In red, we show the bucklebot equilibrium width $w/\ell$. The region below (resp. above) $w/\ell$ indicates slits smaller (resp. larger) than the equilibrium width.  All the trials above this curve have a $100\%$ chance of passing (we send our robots straight onto the slit). However, a sizable region below the curve also sees significant success. We rationalize this region boundary of such success by considering the minimal length the microbots can bend the structure, i.e., $\pi\sqrt{B/F}$, which coincides with the width of the smallest slits that bucklebots can pass. 

%This factor significantly enhances the bucklebot's ability to navigate through constrictions. This deformability is particularly pronounced in the high probability area for large $F\ell^2/B$ and small gaps, enabling effective passage through such configurations. We model this situation again as a hinge-hinge boundary condition problem of a beam of length that matches the gap size ($\delta$), for which the Euler’s critical load is given by $F\delta^2/B=\pi^2$. The corresponding black curve in Fig \ref{fig3}(c) introduces a new division in the phase diagram, which captures the enhanced probability of passing the gap due to its deformability.

\begin{figure}[!ht]

    \includegraphics[width=1\textwidth]{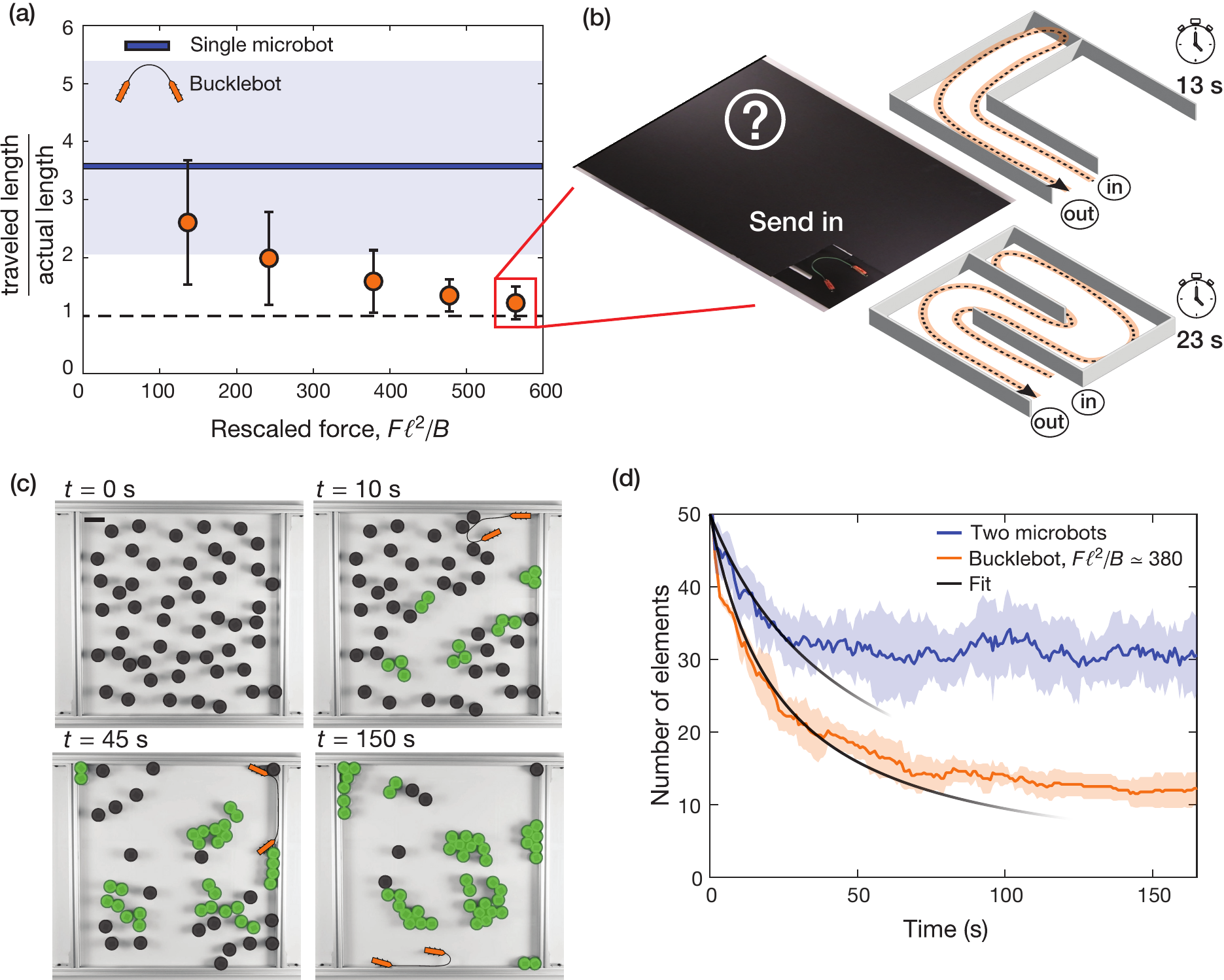}
    \caption{\textbf{Bucklebots probing a closed path and "storing" a room:} (a) Traveled length over actual path length is plotted for bucklebots with a wide range of $F\ell^2/B$. Error bars show the standard deviation of bucklebots' traveled lengths. The solid blue line shows the benchmark for a single microbot and the shaded blue area is its error range. (b) The left snapshot shows a probing experiment: a bucklebot with $F\ell^2/B \simeq 560$ is sent into a covered closed path. The schematic drawings show two paths (longer/shorter) the bucklebots can probe and differentiate. In 14 and 25 seconds, the bucklebot reappears at the starting point of the shorter and longer path, respectively. (c) Snapshots of the evolution of a confined room stored by the same bucklebot. The black circles denote isolated obstacles and the green boundaries correspond to formed clusters. (d) The number of elements representing single or connected obstacles is plotted against time in the case of two single microbots and a bucklebot with $F\ell^2/B \simeq 380$. Each shaded area denotes the standard deviation within 5 trials. Two black lines are the fit derived from the coagulation theory.}
    \label{fig4}
\end{figure}

\section{Discussion}
To summarize, our bucklebots, consisting of two self-propelled microbots coupled by a soft elastic beam, achieve persistent ballistic motions, follow walls, and squeeze their deformable structures through narrow constrictions. The combination of these unique capabilities allows them to perform tasks that individual microbots cannot achieve, such as solving a maze (Fig. \ref{fig1}(c)). In the remaining, we leverage these emergent abilities and demonstrate that the bucklebots can accomplish a broad range of tasks. 

When sent into a closed path, a bucklebot will navigate to the closed end, bounce back,  and reappear at the starting point (see Movie S4). In Fig. \ref{fig4}(a), we show that the ratio between the length traveled by the robots rescaled by the length of the path. While individual microbots travel on average nearly 4 times more than necessary (with nearly 100$\%$ variability between trials), we find that our bucklebots converge to the optimal path as $F\ell^2/B$ increases (while dramatically reducing variability). 
In this limit, our bucklebots can be used to probe and classify simple structures (see Fig. \ref{fig4}(b), where the identification is achieved by recording the entry and exit times).

Likewise, bucklebots differ from the behavior of individual microbots when interacting with obstacles they can displace. In Fig. \ref{fig4}(c), we report a few snapshots of a bucklebot confined with  initially dispersed cylindrical obstacles ($N_0=50$). We find that the bucklebot ($F\ell^2/B\simeq380$) pushes the light obstacles and assembles them into clusters. The number of elements saturates in about a minute. In Fig. \ref{fig4}(d), we report the dynamics of cluster formation for this bucklebot and contrast it with the situation where two microbots freely travel into a similar enclosure. In both cases, we observe an initial decrease in dispersed elements, $N$, before reaching saturation. Bucklebots store nearly $76 \%$ of obstacles into clusters, while two microbots only store $38\%$ of them.
%much more effectively than the single microbots, as suggested by the comparison of storage efficiency $1-N_{\infty}/N_0=76 \%$ for a bucklebot against $38\%$ for a pair of single agents. %as the final number of dispersed elements in its case is significantly lower than that of single microbots. 

Further differences arise when fitting the data with a Smoluchowski-like equation for coagulation\cite{friedlander2000smoke}, $N(t)=N_0/(1+t/\tau)$. The corresponding coagulation time scale $\tau$ indicates a faster decay for the bucklebot ($\tau=23.3 s$) than for single agents ($\tau=49 s$). Additionally, bucklebot interacts more gently with the clusters than single microbots, preventing damage and thus facilitating the formation of larger clusters. The distance  between these assemblies is about $w$, the bucklebot width (Fig.~\ref{fig4}c).

We have shown that stochastic self-propelled active particles coupled with nonlinear elasticity can be tamed and forced into ballistic motion and display various emergent abilities as they interact with different boundaries. These autonomous elasto-active structures carry out all these tasks without directed control. Instead, our elastic model can rationalize and capture these behaviors. We have demonstrated that this newly gained understanding can be leveraged to achieve and control complex tasks, such as maze navigation, probing the length of a path, and collecting cylinders.  Our work on these elasto-active structures thus opens a new pathway for designing soft robotic systems that can adjust and adapt to their surroundings without human intervention.

\section*{Acknowledgments}
It is a pleasure to acknowledge Antoine Deblais and Thomas Barois for helpful discussions, as well as funding from NSF through grants NSF CMMI 2343539, CMMI FMRG 2037097, and financial support from the Princeton High Meadows Environmental Institute.

\section{Materials and Methods}
\subsection{Bucklebot design and manufacturing}
Our active agents are commercially available battery-powered vibrating microbots (Hexbug Nano). Each microbot has a length of 45 mm, a width of 15 mm, a height of 15 mm, and a mass of 7.5 g. Its motion is generated from an internal vibration of a rotating motor transmitted to 12 soft rubber legs to achieve a speed of approximately $154 \pm15$ mm/s. 

The beams are cut from shim stocks using a laser cutter (Epilog Helix-60 Laser engraver). The shim stocks are made of polyester with an elastic modulus of 2 GPa. The thickness and length of such elastic beams are well calibrated to ensure a variation of bending stiffness used in experiments.

The collar that is used to connect microbots with elastic beams is designed by Rhino and 3D printed by Prusa i3 printer using poly-lactic acid (PLA) (density $\rho$ = 1.2 g/cm\textsuperscript{3} and elastic modulus $E$ = 5 GPa). The beams are clamped to the collars using Dodge 0-80 .115 inch length inserts and corresponding screws.

\subsection{Experimental setups and bucklebot tracking}
The active force exerted by the microbot is estimated by measuring its pushing force via an Instron 10N load cell. The active force is measured to be $20 \pm3$ mN. We choose the microbot pairs with approximately the same free velocity and active force to ensure experiment consistency. It is worth noting that the microbot's manufacturing defects and component variabilities give rise to its biased motion. Experimentally, a biased microbot performs a circular motion, whose radius is given by $R=v_f/\omega_b$, where $\omega_b$ is the angular rotation rate. We adopt the criteria from Baconnier et al. \cite{baconnier:tel-04081179} and choose the microbots that are not noticeably biased. All experiments are carried out on an acrylic surface. For bucklebots, we change the two microbots' batteries simultaneously to maintain their same relative battery level throughout the experiments. 

To capture the motion of the microbots and the bucklebots, a Canon EOS 80D camera is held by a frame looking down at a large white cast acrylic sheet from McMaster-Carr on top of the lab table. To track these robots while effectively differentiating each individual from one another, we use binary square fiducial markers, known as ArUco markers, which are synthetic square markers composed of a wide black border and an inner binary matrix that determines its identifier (id). We print out markers with different IDs and attach them to each microbot present in the experiments. With Python’s Open Source Computer Vision (OpenCV) package\cite{opencv_library}, we post-process the recorded videos by tracking the attached markers' position data ($x,y,t$) with time. For example, our code detects the position ($x,y$) of the marker's four corners. We calculate the mid-point positions of opposite edges on each marker, which allows us to obtain the orientation vector of the microbots. In addition, the velocity of a single microbot is measured by multiplying its position displacement of consecutive frames with frames per second (fps), which allows us to further calculate the mean velocity by averaging the marker's velocities over time. We estimate a bucklebot's center of mass position as the line's center point that connects the two marker centers. 

\section{Bucklebot model}
\begin{figure}[!ht]
    \centering
    \includegraphics[width=1.0\textwidth]{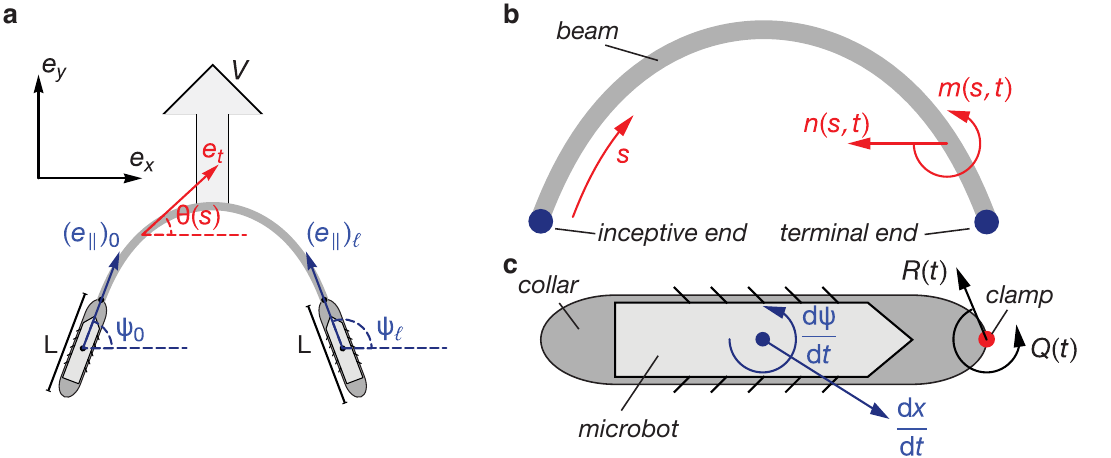}
    \caption{\textbf{bucklebot model schematic}
    (\textbf{a}) bucklebot traveling in the $y$-direction.
    (\textbf{b}) Schematic of a beam.
    (\textbf{c}) Shematic of a microbot with a collar that clamps at the front.
    }
    \label{fig:bucklebot-schematic}
\end{figure}

 We describe an analytical model of the bucklebot that couples the beam dynamics and the microbot's self-propelled motion. We begin by introducing the beam equations and self-propelled microbot equations, rescaling the beam and microbot equations by common length, time, and force scales, and discussing the four dimensionless groups that describe the general dynamics. We then discuss the mathematical constraints imposed by the collar that clamps the beam to the microbots in the bucklebot configuration. Next, we justify and introduce the time-stepping algorithm used to simulate the bucklebots. Finally, we derive the analytic results that provide predictions for the bucklebot velocity, shape, and onset of buckling. 
\subsection{Equations and rescaling}
The beam dynamics can be described by the 2D Kirchhoff equations
    \begin{subequations}
    \begin{align}
        \frac{\partial \bm{n}}{\partial s}
        &=
        \rho_{\text{b}} \frac{\partial^2 \bm{r}}{\partial t^2},  \label{eq:beam_force} \\
        \frac{\partial m}{\partial s} + \bm{e}_{\text{t}} \times \bm{n}
        &= 0,
        \label{eq:beam_mom}
    \end{align}\label{eq:Kirchhoff}
 with the constitutive equation
    \begin{align}
        m &= B \frac{\partial \theta}{\partial s}. \label{eq:beam_con}
    \end{align} \label{eqs:beam}
    \end{subequations}
Here $\bm{n} \left( s,t\right)$ is the internal force, $m\left(s,t\right)$ is the internal moment, $s \in \left[ 0 , \ell \right]$ is the arc-length position along the beam, $t$ is time, $\bm{r}\left(s,t\right)$ is the center-line position, and $\tfrac{\partial \bm{r}}{\partial s}{=}\bm{e}_{\text{t}}\left(s,t\right){=}\left\{ \cos \theta ,\sin \theta \right\}$ is the unit tangent with $\theta\left(s,t\right)$ the angle between the beam tangent $\bm{e}_{\text{t}}$ and $x$-axis $\bm{e}_x$. $\rho_{\text{b}}$, and $B$ are material parameters representing the beam linear density and bending stiffness, respectively.

The self-propelled motion of the microbots obeys the force and moment balance:
\begin{subequations}
    \begin{align}
        M \frac{\diff^2{\bm{x}}}{\diff{t}^2} % \frac{\partial^2 \bm{x}_i}{\partial t^2} 
        + \gamma \frac{\diff{\bm{x}}}{\diff{t}} %\frac{\partial \bm{x}_i }{\partial t} 
        &=
        F \bm{e}_{\parallel} 
        + \bm{R}
        ,\label{eq:hex_force} \\
        I \frac{\diff^2{\psi}}{\diff{t}^2} %\frac{\partial^2 \varphi_i }{\partial t^2} 
        + \Gamma \frac{\diff{\psi}}{\diff{t}} %\frac{\partial \bm{\varphi}_i}{\partial t} 
        &=
        \frac{1}{2}L \left( \bm{e}_{\parallel} \times \bm{R} \right)
        + Q 
        .\label{eq:hex_mom}
    \end{align}
\label{eqs:hex}
\end{subequations}
where $\bm{x}\left(t\right)$ is the center-of-mass, $\bm{e}_{\parallel}\left(t\right)=\left\{ \cos \psi, \sin \psi \right\}$ is the unit orientation vector aligned along the long-axis of the microbot, and $\psi\left(t\right)$ the angle of the orientation vector $\bm{e}_{\parallel}$ with $x$-axis $\bm{e}_x$. $M$, $I$, $F$, $\gamma$, $\Gamma$, and $L$ are material and geometric parameters of the microbot representing the mass, moment of inertia, driving force, translational damping coefficient, rotational damping coefficient, and length respectively. $\bm{R}\left( t \right)$ and $Q\left( t \right)$ are the reaction force and reaction torque acting on the microbots with $\tfrac{1}{2} L \left( \bm{e}_{\parallel} \times \bm{R} \right)$ being the moment of force from the reaction force being applied on the collar away from the microbot center of mass. 

The bending stiffness $B$, mass $M$, driving force $F$, translational friction coefficient $\gamma$, beam length $\ell$, and microbot (plus collar) length $L$ are measured in the lab as described in the Materials and Methods section. The linear beam density $\rho_{\text{b}}$ is taken from manufacturer data and beam geometry. The moment of inertia $I=\iint_{\mathcal{R}} \rho_{\text{H}}r^2 \diff{A}$ and the rotational damping coefficient $\Gamma=\iint_{\mathcal{R}} \rho_{\mathrm{\gamma}} r^2 \diff{A}$ are derived quantities from the microbots area mass density $\rho_{\text{H}}$ and damping density $\rho_{\mathrm{\gamma}}$ as well as the distance from the center of mass $r$ over the microbot body $\mathcal{R}$. For simplicity, we assume a rod-shaped body $W \ll L$ with uniform densities $\rho_{\text{H}}=\tfrac{M}{LW}$ and $\rho_{\mathrm{\gamma}}=\tfrac{\gamma}{LW}$ such that $I \approx \tfrac{1}{12}ML^2$ and $\Gamma \approx \tfrac{1}{12}\gamma L^2$.

Rescaling the length by the beam length $\left\{\bm{r}, \bm{x}, s \right\}{=}\left\{\bm{r} / \ell, \bm{x} / \ell, s / \ell \right\}$, the time by $\{t\}{=} \{t / \sqrt{M \ell^3 / B}\} $, the force by $\left\{ \mathcal{F}, \bm{R} \right\}{=}\left\{ F \ell^2 / B, \bm{R} \ell^2 / B \right\}$, the moment by $\left\{ m ,Q \right\} {=} \left\{ m \ell/B , Q \ell / B \right\}$, and taking $I=\dfrac{1}{12}M L^2$ and $\Gamma = \tfrac{1}{12}\gamma L^2$, we arrive at the dimensionless equations for the beam:
\begin{subequations}
    \begin{align}
        \frac{\partial \bm{n}}{\partial s}
            &=
            \mathcal{M} \frac{\partial^2 \bm{r}}{\partial t^2}
            ,\label{eq:beam_force_non} \\
        \frac{\partial m}{\partial s} + \bm{e}_{\text{t}} \times \bm{n}
            &= 0
            ,\label{eq:beam_mom_non} \\
        m &= \frac{\partial \theta}{\partial s} ,
    \end{align}\label{eqs:non_beam}
\end{subequations}
and the microbots:
\begin{subequations}
    \begin{align}
        \frac{\diff^2{\bm{x}}}{\diff{t}^2} %\frac{\partial^2 \bm{x}}{\partial t^2} 
            + \zeta  \frac{\diff{\bm{x}}}{\diff{t}} % \frac{\partial \bm{x} }{\partial t} 
            &=
            \mathcal{F} \bm{e}_{\parallel} 
            + \bm{R}
            ,\label{eq:hex_force_Pi} \\
        \frac{\mathcal{L}^2}{12} \left( \frac{\diff^2{\psi}}{\diff{t}^2} %\frac{\partial^2 \varphi }{\partial t^2} 
            +  \zeta \frac{\diff{\psi}}{\diff{t}} \right)%\frac{\partial \varphi}{\partial t}
            &=
            \frac{\mathcal{L}}{2} \left( \bm{e}_{\parallel}
            \times \bm{R} \right)
            + Q
            .\label{eq:hex_mom_Pi}
    \end{align}
    \label{eqs:non_hex}
\end{subequations}
The rescaled and rearranged Eqs.~\ref{eqs:non_beam}-\ref{eqs:non_hex} introduce four dimensionless groups: $\mathcal{M}=\rho_{\text{b}} \ell/ M $, $\mathcal{L}=L/\ell$, $\mathcal{F}=F\ell^2/B$, and $\zeta=\gamma/\sqrt{MB/\ell^3}$. $\mathcal{M}$ and $\mathcal{L}$ are self-explanatory as they compare the mass and lengths, respectively. In our system, the microbot is always much heavier than the beam such that $\mathcal{M} \ll 1$, and consequently, the beam dynamics are quasi-static. If the microbot were to have zero length $\mathcal{L}=0$, the torque balance (Eqn.~\ref{eq:hex_mom_Pi}) would simplify to $Q=0$ such that the microbots would not resist moments. The microbots, in this case, could be considered point particles with an orientation-dependent driving force. $\mathcal{F}$ represents an elasto-active number that compares the driving force of the microbot to the beam's resistance to deformation. When $\mathcal{F}$ is large, the microbot easily deforms the beam, and when $\mathcal{F}$ is small, the beam remains undeformed. The dimensionless group $\zeta$ is the damping ratio of the microbot with a spring constant $B/\ell^3$. In the actual range of parameters tested, $\zeta \gg 1$ so that we may assume overdamped microbot dynamics. 
% The final simplified equations are:
%     \begin{align}
%         \frac{\partial \boldsymbol{n}}{\partial s}
%             &=
%             0
%             ,\label{eq:beam_force_non} \\
%         \frac{\partial m}{\partial s} + \bm{e}_{\text{t}} \times \bm{n}
%             &= 0
%             ,\label{eq:beam_mom_non} \\
%         m &= \frac{\partial \theta}{\partial s},
%     \end{align}
% \begin{align}
%         \frac{\diff{\bm{x}}}{\diff{t}} % \frac{\partial \bm{x} }{\partial t} 
%             &=
%             \mathcal{F} \bm{e}_{\parallel} 
%             + \bm{R}
%             ,\label{eq:hex_force_Pi} \\
%         \frac{\lambda^2}{12}  \frac{\diff{\psi}}{\diff{t}} %\frac{\partial \varphi}{\partial t}
%             &=
%             \frac{\lambda}{2} \left( \bm{e}_{\parallel} 
%             \times \bm{R} \right)
%             +Q
%             .\label{eq:hex_mom_Pi}
%     \end{align}

\subsection{Bucklebot constraints}
The governing equations for the beam (Eqs.~\ref{eqs:beam}) are coupled to the governing equations for the microbots (Eqs.~\ref{eqs:hex}) due to the clamping between the beam ends and the microbot collars. The beam is clamped to two microbots at its inceptive $(s=0)$ and terminal $(s=\ell)$ ends as shown in Fig.~\ref{fig:bucklebot-schematic}. To distinguish the two microbots we use the subscripts $\left(\cdot\right)_0$ and $\left(\cdot\right)_\ell$ to represent the inceptive and terminal end microbots, respectively. The relations summarize the coupling:
\begin{subequations}
\begin{align}
    & \bm{r}\left(s{=}0\right) = \bm{x}_0+\frac{L}{2} \left(\bm{e}_{\parallel}\right)_0,& 
    & \bm{r}\left(s{=}\ell\right) = \bm{x}_{\ell}+\frac{L}{2} \left(\bm{e}_{\parallel}\right)_{\ell},\\
    & \underbrace{
        \bm{e}_{\text{t}}\left(s{=}0\right) = \left(\bm{e}_{\parallel}\right)_0 }_{
        \text{\emph{i.e.,} }
        \theta = \psi_0}
        ,& 
    & \underbrace{
        \bm{e}_{\text{t}}\left(s{=}\ell\right) = -\left(\bm{e}_{\parallel}\right)_{\ell} }_{
        \text{\emph{i.e., }}\theta = \psi_{\ell} - \pi}
        ,\\
    & \bm{n}\left(s{=}0\right) = \bm{R}_0,&
    & \bm{n}\left(s{=}\ell\right) = -\bm{R}_{\ell},\\    
    & m\left(s{=}0\right) = Q_0 ,&
    & m\left(s{=}\ell\right) = -Q_{\ell} . 
\end{align}\label{eq:constraints}
\end{subequations}
Or, if we recast Eqs.~\ref{eq:constraints} in a dimensionless form consistent with Eqs.~\ref{eqs:non_beam}--\ref{eqs:non_hex} and where the terminal end microbot variables are denoted with the subscript $\left(\cdot\right)_1$:
\begin{subequations}
\begin{align}
    & \bm{r}\left(s{=}0\right) = \bm{x}_0+\frac{\lambda}{2} \left(\bm{e}_{\parallel}\right)_0,& 
    & \bm{r}\left(s{=}\ell\right) = \bm{x}_{1}+\frac{\lambda}{2}\left(\bm{e}_{\parallel}\right)_{1},\\
    & \underbrace{
        \bm{e}_{\text{t}}\left(s{=}0\right) = \left(\bm{e}_{\parallel}\right)_0 }_{
        \text{\emph{i.e.,} }
        \theta = \psi_0}
        ,& 
    & \underbrace{
        \bm{e}_{\text{t}}\left(s{=}1\right) = -\left(\bm{e}_{\parallel}\right)_{1} }_{
        \text{\emph{i.e., }}\theta = \psi_{1} - \pi}
        ,\\
    & \bm{n}\left(s{=}0\right) = \bm{R}_0,&
    & \bm{n}\left(s{=}1\right) = -\bm{R}_{1},\\    
    & m\left(s{=}0\right) = Q_0 ,&
    & m\left(s{=}1\right) = -Q_{1} . 
\end{align}\label{eq:constraints_non}
\end{subequations}

\subsection{Time-stepping algorithm}

Eqs. \ref{eq:beam_force_non}--\ref{eq:hex_mom_Pi} are solved using a time-stepping algorithm with initial and boundary conditions in Mathematica to generate Fig. 2(a) and compare with experiments. Specifically, we start with two microbots at particular orientations whose fronts face each other. We take them as boundary conditions and use a numerical shooting method to solve for the beam's moment and force, $\boldsymbol{n}$ and $\boldsymbol{m}$, respectively. With the beam moment and force matching $\boldsymbol{R}$ and $\boldsymbol{Q}$, we calculate the translational and rotational accelerations using Eqs. \ref{eq:hex_force_Pi}-\ref{eq:hex_mom_Pi}. We then multiply accelerations over $\Delta t$ to get new velocities and velocities over $\Delta t$ to get new positions and orientations. And we are at the next time step with new positions and orientations to solve for the beam shape, moment,  and force.

\subsection{Steady-state solutions for bucklebots}

To capture the bucklebots' final shape and velocity, we solve the beam shape and the steady-state translation of the microbots simultaneously. For the bucklebot motion, we assume a purely transverse motion in the $\bm{e}_y$ direction so that traveling velocity is $\bm{v} {=} v_y \bm{e}_y $. Here we assume the microbots have equal velocity $\bm{v}{=}\bm{v}_1{=}\bm{v}_2$, zero angular velocity $\omega{=}0$. We assume the beam is quasistatic $\dfrac{\partial\bm{n}}{\partial s} {=} 0$ such that the reaction moments are equal $\bm{R}{=}\bm{R}_0{=}\bm{R}_1$. Further, we assume reflective symmetry about the transverse velocity direction. Therefore, the orientation angles of the microbots may be described as $\psi{=}\psi_0{=}\pi{-}\psi_1$. As a result, the force and moment balances (eqs. \ref{eqs:hex}) for the two attached microbots of a bucklebot at steady-state are simplified as:
\begin{subequations}
\begin{align}
    0 &= \mathcal{F}\cos\psi+R_x,\label{hexleft_sseq_vx} \\
    v_y &= \mathcal{F}\sin\psi+R_y,\label{hexleft_sseq_vy} \\
    0 &= \frac{1}{2}\lambda(R_y\cos\psi-R_x\sin\psi)+Q_0, \label{hexleft_sseq_m}
\end{align}\label{eq:hexleft}
\end{subequations}
\begin{subequations}
\begin{align}
    0 &= -\mathcal{F}\cos\psi-R_x,\label{hexright_sseq_vx}\\
    v_y &= \mathcal{F}\sin\psi-R_y,\label{hexright_sseq_vy}\\
    0 &= \frac{1}{2}\lambda(R_y\cos\psi+R_x\sin\psi)-Q_1\label{hexright_sseq_m},
\end{align}\label{eq:hexright}
\end{subequations}
where eqs. \ref{eq:hexleft} describe the microbot attached at the inceptive end and eqs. \ref{eq:hexright} describe the microbot at the terminal end of the beam. Note the sign differences result from the way we define the hexbug angle symmetry $\psi$: $\cos(\pi-\psi)=-\cos\psi$. Solving Eqs. \ref{eq:hexleft}-\ref{eq:hexright} we get:
\begin{subequations}
\begin{align}
    % v_x &= 0,\label{ss_vx}\\
    v_y &= \mathcal{F}\sin\psi,\label{ss_vy}\\
    R_x &= -\mathcal{F}\cos\psi,\label{ss_rx}\\
    R_y &= 0,\label{ss_ry}\\
    Q_0 &= -\frac{1}{2}\mathcal{F}\lambda\cos\psi\sin\psi.\label{ss_q0} \\
    Q_1 &= -\frac{1}{2}\mathcal{F}\lambda\cos\psi\sin\psi.\label{ss_ql}
\end{align}    
\end{subequations}
Finally, we match the beam boundary conditions:
\begin{align}
    &\boldsymbol{n}(0)=\boldsymbol{n}(1) = -\mathcal{F}\cos\psi \boldsymbol{e}_x,\\
    &m(0)=m(1)=-\frac{1}{2}\mathcal{F}\lambda\cos\psi\sin\psi,\\
    &y(0)=y(1)=0\\
    &\theta(0)=-\theta(1)=\psi.
\end{align}
 and solve Eqs. \ref{eq:beam_force_non}-\ref{eq:beam_mom_non} to retrieve the steady-state shape ($\theta(s)$, $x(s)$, $y(s)$) of the beam. Plugging $\psi=\theta(0)$ in Eqn. \ref{ss_vy}, we get the steady-state velocity of the bucklebot. The equilibrium width of the bucklebot, which appears in Fig. \ref{fig3}(d), is defined as $x(1)-x(0)$.

\subsection{The relative length of the microbot to beam affects the onset of buckling}

\begin{figure}[!ht]
    \centering
    \includegraphics[width=1.0\textwidth]{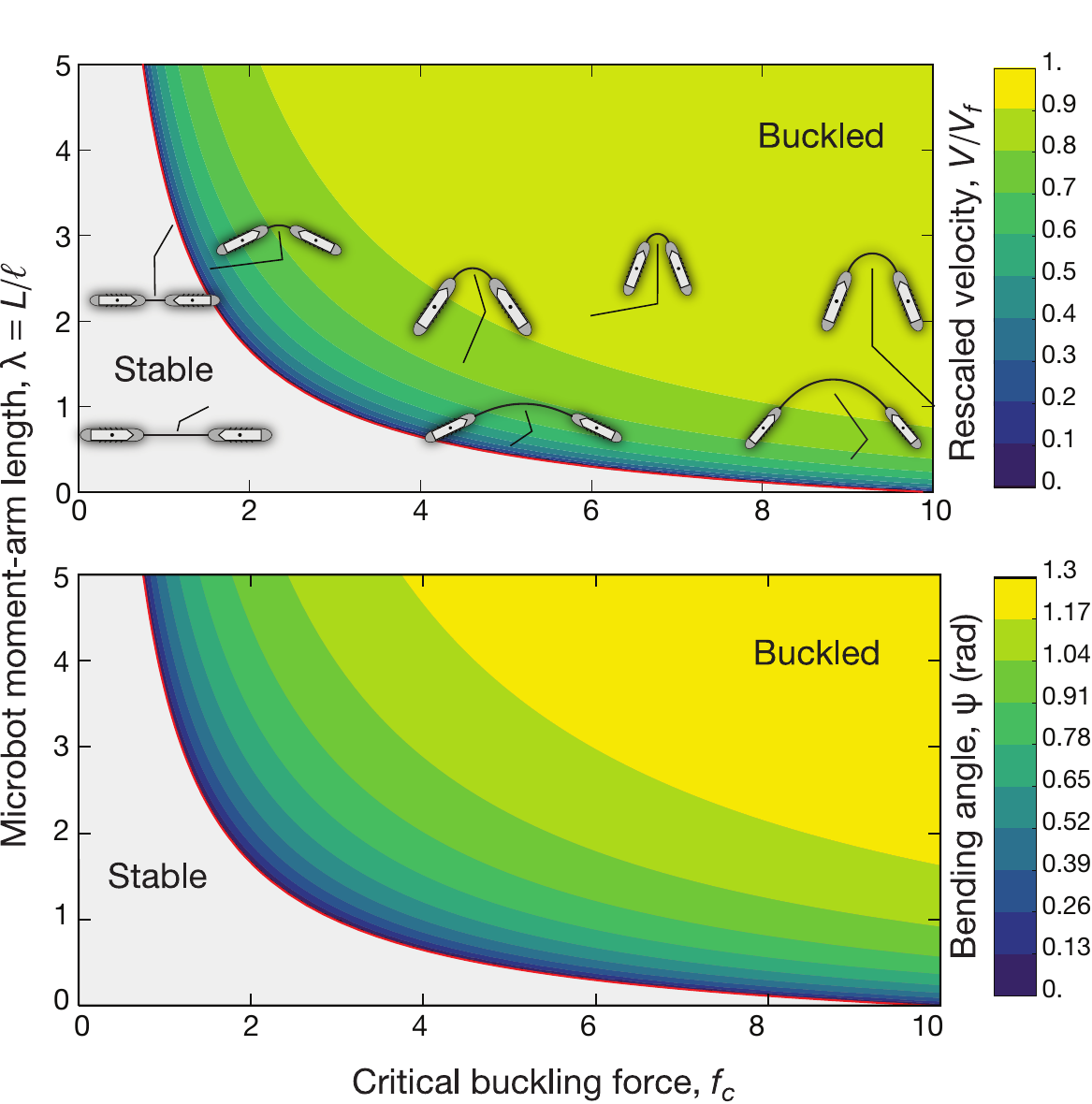}
    \caption{\textbf{Bucklebot state diagram}  
    The beam buckles and causes a directed velocity (top) and bending angle $
    \psi$ (bottom) according to the rescaled force and the relative microbot-beam length $\lambda$. The red lines correspond to Eqn. \ref{eq:critical force}, and the areas above the line are color-coded by the steady-state solutions described in Section D.
    }
    \label{fig:bucklebot-buckling}
\end{figure}

In Fig. 2(b)-(c)'s theory curves, we plot the limit case $\lambda\simeq 0$ where microbots are regarded as point masses. In experiments, however, this is not the most accurate case, and we introduced the third dimensionless number $\lambda=L/\ell$. Here, we further discuss how $\lambda$ affects the behavior of bucklebots at the onset of buckling. 
Eqn. \ref{eq:beam_mom_non} can be expanded as:
\begin{align}
    m'+n_x\sin\theta-n_y\cos\theta=0
\end{align}
At the onset of buckling, we take $n_y\simeq0$ and $n_x=\mathcal{F}\cos\theta$, so that the above equation can be rewritten as: 
\begin{align}
    m'+\mu^2y'=0 \Rightarrow m+\mu^2y=m_0
\end{align}
where $\mu=\sqrt{\mathcal{F}\cos\theta}$ and $m_0=-\frac{1}{2}\mathcal{F}\lambda\cos\theta\sin\theta$. Taking $m=\theta'$ and linearizing about small $\theta \ll 1$ such that $\theta\simeq dy/dx$ we get an ordinary second-order differential equation with boundary conditions: 
\begin{align}
    &y''+\mu^2y=m_0,\\
    &y(0)=0 \ \text{and}\ y(1)=0.
\end{align}
with the solution
\begin{align}
    y(x)=-\frac{m_0(-1+\cos(\mu x)+\sin(\mu x)\tan(\mu/2))}{\mu^2}
\end{align}
We seek solutions where the microbots rotate at a very small angle $\psi=\theta_0$
\begin{align}
    \theta_0\simeq y'(0)=\frac{m_0\tan(\mu/2)}{\mu}
\end{align}
Plugging in the definition of $\mu$ and $m_0$ and linearizing about small $\theta_0$ with a Taylor expansion we get
\begin{align}
   \theta_0=\frac{1}{2}\lambda\sqrt{\mathcal{F}}\tan(\sqrt{\mathcal{F}}/2)\theta_0+O(\theta_0^2)
\end{align}
Therefore, we have an implicit function for the critical force of buckling $\mathcal{F}=f_c$, which we plot as the red line in Fig. \ref{fig:bucklebot-buckling}:
\begin{align}
    \lambda=\frac{2\cot(\sqrt{f_c/2})}{\sqrt{f_c}} \label{eq:critical force}
\end{align}

\subsection{Euler's critical load}
According to our theory curve in Fig. \ref{fig2}(b)-(c) and prediction from Eqn. \ref{eq:critical force} , in the limit case of $\lambda \simeq0$, the critical force of buckling $F_c\ell^2/B\simeq \pi^2$. Here, we mathematically verify this result. 

Consider a slender column with length $\ell$ supported at each hinged end with axial forces $F$ applied at each end. A summation of moments about point $x$ along the curve yields:
\begin{align}
    \sum M=0 \Rightarrow M(x)+Fw=0
\end{align}
where $w$ is the lateral deflection. According to Euler-Bernoulli beam theory, the deflection of the beam can be related to its bending moment by $M=-B\frac{d^2w}{dx^2}$ so that:
\begin{align}
    \frac{d^2w}{dx^2}+\mu^2w=0
\end{align}
where $\mu^2=\frac{F}{B}$. One can solve this ordinary differential equation with boundary conditions $w(0)=w(\ell)=0$ and yield that $\mu_n=\frac{n\pi}{\ell}$, for $n\in \mathbb{N}$. Therefore,
\begin{align}
    F_n=\frac{n^2\pi^2B}{\ell^2}, \ \text{for}\ n\in \mathbb{N}
\end{align}
Theoretically, the column is more prone to buckle to its first mode because of lower energy \cite{audoly2000elasticity} so that
\begin{align}
    F_{c}=\frac{\pi^2 B}{\ell^2}\Rightarrow F_c\ell^2/(B)=\pi^2
    \label{eq:hinge-hinge}
\end{align}
To generate the solid black line in Fig. \ref{fig3}(d), we refer to this derivation to demonstrate the relationship between the bucklebot rescaled force and the slit width $\delta$. From Eqn. \ref{eq:hinge-hinge},  the minimal length the microbots can bend the beam $\ell_{min}=\pi \sqrt{B/F}$, which should match the width of the smallest slit that the microbot can pass. Therefore,
\begin{align}
    \delta=\pi \sqrt{B/F} \Rightarrow F\delta^2/B=\pi^2 
    \label{eq:hinge-hinge2}
\end{align}
Multiplying both sides of  Eqn. \ref{eq:hinge-hinge2} by $\ell^2$ and rearranging we can arrive at
\begin{align}
    \delta/\ell= \pi^2B/F\ell^2 \label{eq:hinge_hinge3}
\end{align}

\subsection{Dynamics of a single tail}
Analogous to the behavior of the outer microbot we observe in Fig. \ref{fig3}(a), we refer to the self-oscillation of a simple configuration where a single microbot is paired with an elastic beam clamped on one end.  This configuration can be dynamically modeled using the same Eqs. \ref{eq:beam_force_non}-\ref{eq:hex_mom_Pi} but with modified boundary conditions: 
\begin{align}
&\boldsymbol{R}=-\boldsymbol{n}(s=\ell)\ \text{and} \ \boldsymbol{Q}=-m(s=\ell)\cdot e_z \ \text{for microbot},\label{eq:tail_bc1}\\
&\boldsymbol{r}(0)=\theta(0)=0 \ \text{for beam at all time}.\label{eq:tail_bc2}
\end{align}
We solve Eqs. \ref{eq:beam_force_non}-\ref{eq:hex_mom_Pi} using a time-stepping algorithm described in Section C with different $F\ell^2/B$ and extract the period of oscillation to generate the solid lines in Fig. \ref{fig3}(b). For example, in the case of $\alpha=\pi/6$, we extract from the modeled dynamics the time it takes for the microbot to oscillate from an initial position $\psi=\pi/6$ to $\psi=\pi$ for different rescaled forces. And we repeat the same procedure for the cases $\alpha=\pi/3$ and $\alpha=4\pi/9$.

\end{document}